# Enhancing Conversion Gain in HEB Mixers using Negative Differential Resistance

Boris S. Karasik, *Senior Member, IEEE*, and Changyun Yoo

***Abstract*—This work investigates how negative differential resistance (NDR) can be utilized to enhance conversion gain in superconducting niobium nitride (NbN) hot-electron bolometer (HEB) mixers. Conventional HEB operation avoids NDR regions due to parasitic oscillations caused by interactions between thermal inertia and bias-line reactance. Recent experimental results show that reducing the bias-T inductance suppresses these oscillations and enables stable biasing within the NDR regime, thereby increasing the mixer conversion gain beyond levels achievable under positive differential resistance (PDR) bias. Using a lumped thermal–electrical model, we interpret these experimental findings, derive stability conditions for NDR-biased operation, and predict the future performance limits achievable through optimized bias circuit designs. Our results indicate that further reductions in bias inductance, combined with appropriate load-line conditions, could yield several decibels of additional gain while maintaining practical IF bandwidths.**

## I. Introduction

INCREASING the conversion gain of superconducting terahertz (THz) hot-electron bolometer (HEB) mixers is highly desirable because it could help narrow the roughly fourfold sensitivity gap between current state-of-the-art noise temperatures and the fundamental limit set by quantum noise and electron-temperature fluctuation–induced thermal noise (Fig. 1) [1]. The earliest work on superconducting HEB mixers [2] recognized that such mixers are not fundamentally constrained by the −3 dB gain limit applicable to resistive mixers. Higher gain becomes possible when the device is biased near the onset of electrothermal instability.

A later analysis [3] suggested that higher conversion gain could be achieved in bias regions exhibiting negative differential resistance (NDR). Historically, however, the presence of parasitic oscillations has prevented systematic exploration of this bias region. As a result, conventional practice has been to avoid the N-shaped portion of the current–voltage characteristic (IVC).

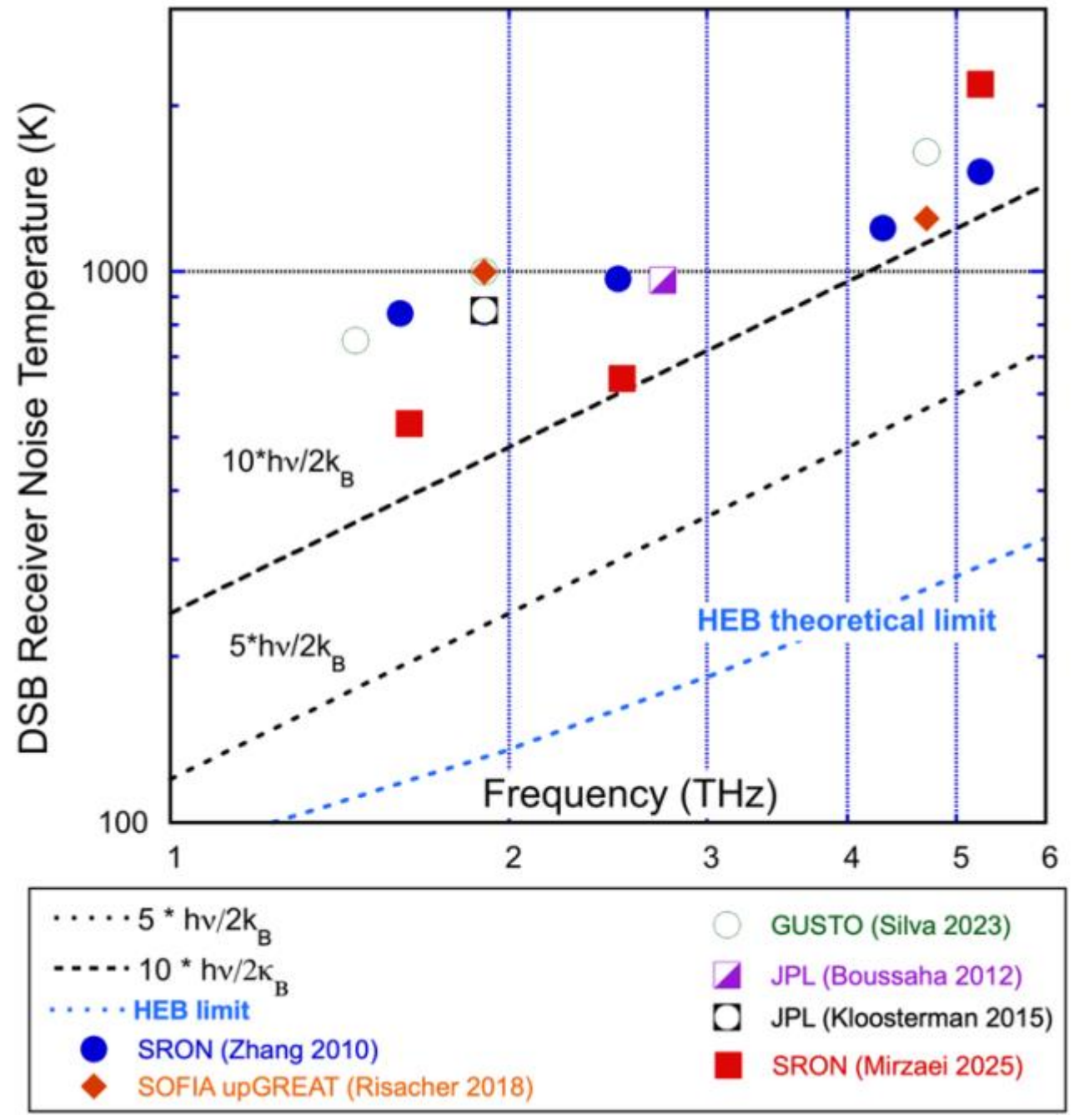

**Fig. 1.** Noise temperature of state-of-the-art THz NbN HEB receivers. The HEB theoretical limit is from [5]. Adapted from [1], with permission from the American Institute of Physics, 2026.

Recent work [1] clarified the origin of these oscillations, identifying them as unsuppressed resonances at frequency $f_{\text{osc}} \approx \left(2\pi\sqrt{C_N L_b}\right)^{-1}$, where $C_N \sim$ a few pF is an effective capacitance (on the order of a few picofarads) associated with the bolometer's thermal inertia, and $L_b$ is the bias-T inductance. Reducing $L_b$ shifts $f_{\text{osc}}$ into a region where dissipation suppresses oscillations, stabilizing the bias point even within the NDR regime.

This raises two key questions: What limits the achievable conversion gain, and what bias conditions enable operation at that limit? Here we use a thermal–electric model to interpret experimental observations in [1] and evaluate the performance potential of future devices.

## II. Summary of the Experimental Results

Experimental verification used a quasi-optical NbN HEB (length 0.2 μm, width 2 μm, thickness 5 nm) integrated with a log-spiral antenna and hyperhemispherical silicon lens. Full

The research described in this paper was carried out at the Jet Propulsion Laboratory, California Institute of Technology, under a contract with the National Aeronautics and Space Administration. C.Y. acknowledges the support from the NASA Postdoctoral Program. *(Corresponding author: Boris S. Karasik).*

Boris S. Karasik, is with the Jet Propulsion Laboratory, California Institute of Technology, Pasadena, CA 91109 USA (email: boris.s.karasik@jpl.nasa.gov).

Changyun Yoo was with the Jet Propulsion Laboratory, California Institute of Technology, Pasadena, CA 91109 USA. He is now with xLight, Palo Alto, CA 94306, USA (e-mail: changyunyoo@gmail.com).

details appear in [1]. For consistency with modeling, essential elements are summarized here.

A diagram of the experimental setup is shown in Fig. 2. The mixer generates a 1.4-GHz IF signal by beating two monochromatic 2.5-THz inputs. This IF signal is first amplified by a low-noise amplifier (LNA) with a noise temperature $T_A \approx$ 2 K. To reduce standing-wave–induced fluctuations in the amplifier noise, a cryogenic L-band isolator is inserted between the mixer and the LNA. The passband of the combination of the LNA and the isolator 0.8-2 GHz. Outside the dewar, the IF output power is further amplified by a room-temperature amplifier chain and measured using a spectrum analyzer.

Figure 3 reproduces the main experimental finding: the conversion gain increases steadily as the bias transitions from the positive differential resistance (PDR) regime into the negative differential resistance (NDR) regime. This transition is driven by decreasing the local-oscillator (LO) power incident on the device. For these measurements, a bias-T inductance of 70 nH was used. As shown in [1], such a low value of $L_b$ was essential for achieving stable operation; however, reducing $L_b$ further comes with a drawback: it would cause excessive leakage of IF power into the bias circuitry.

The resulting IVC in the stable-bias regime displayed a moderately pronounced N-shape. The maximum value of the self-heating parameter $\mathcal{L}$ (aka the electro-thermal feedback (ETF) loop gain) reached 1.15. Its relationship to the DC differential resistance $Z(0) = dV/dI$ is:

$$Z(0) = R\,(1+\mathcal{L})/(1-\mathcal{L}), \tag{1}$$

where $R = V/I$ is the ohmic resistance. Equation (1) follows directly from the standard bolometric model [4, 5], which assumes a single, uniform electron temperature across the device (the lumped-element approximation) and a temperature-dependent resistance.

Conversion gain $\eta$ increases steadily as the bias approaches the NDR regime, exceeding the peak PDR gain by several decibels. The maximum PDR gain observed here matches the typical values obtained in other experimental setups employing larger, conventional bias-T inductors. Bias operation in the PDR regime is always stable because the IVC is a single-valued, concave function.

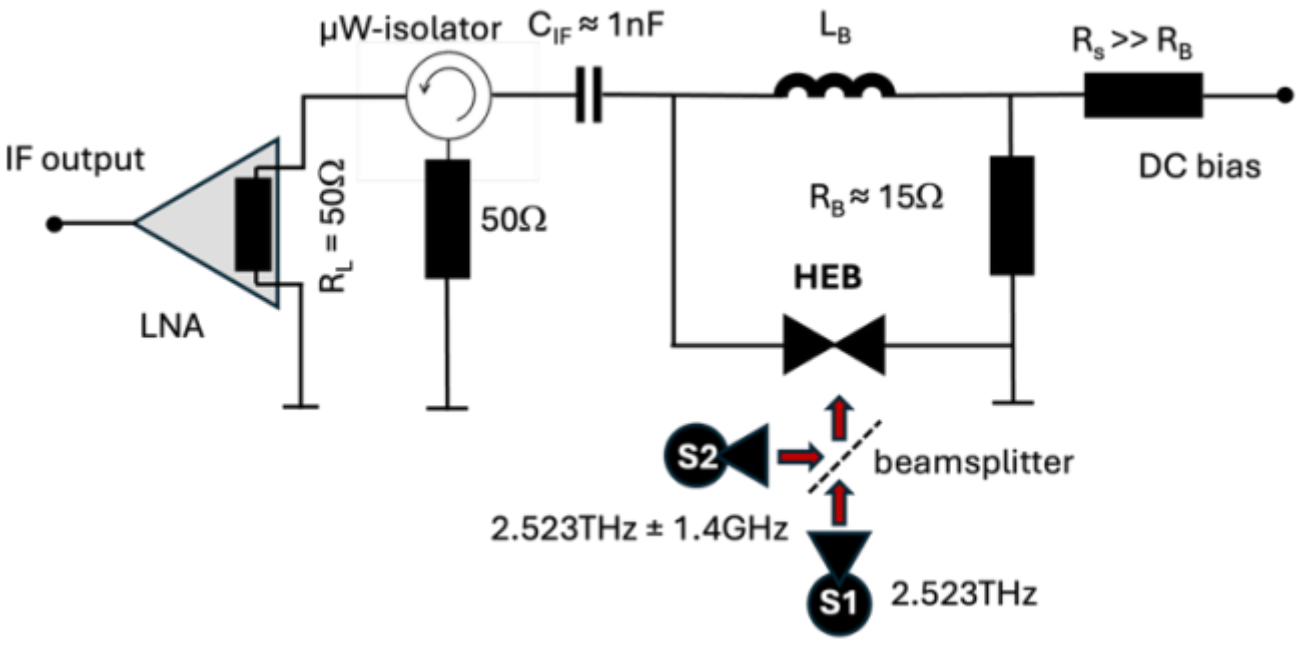

**Fig. 2.** Experimental setup schematic. All the components are at 4.2 K except for the THz sources S1 and S2, which are at room temperature. S1 is a far-IR gas laser, whereas S2 is a tunable frequency multiplier chain.

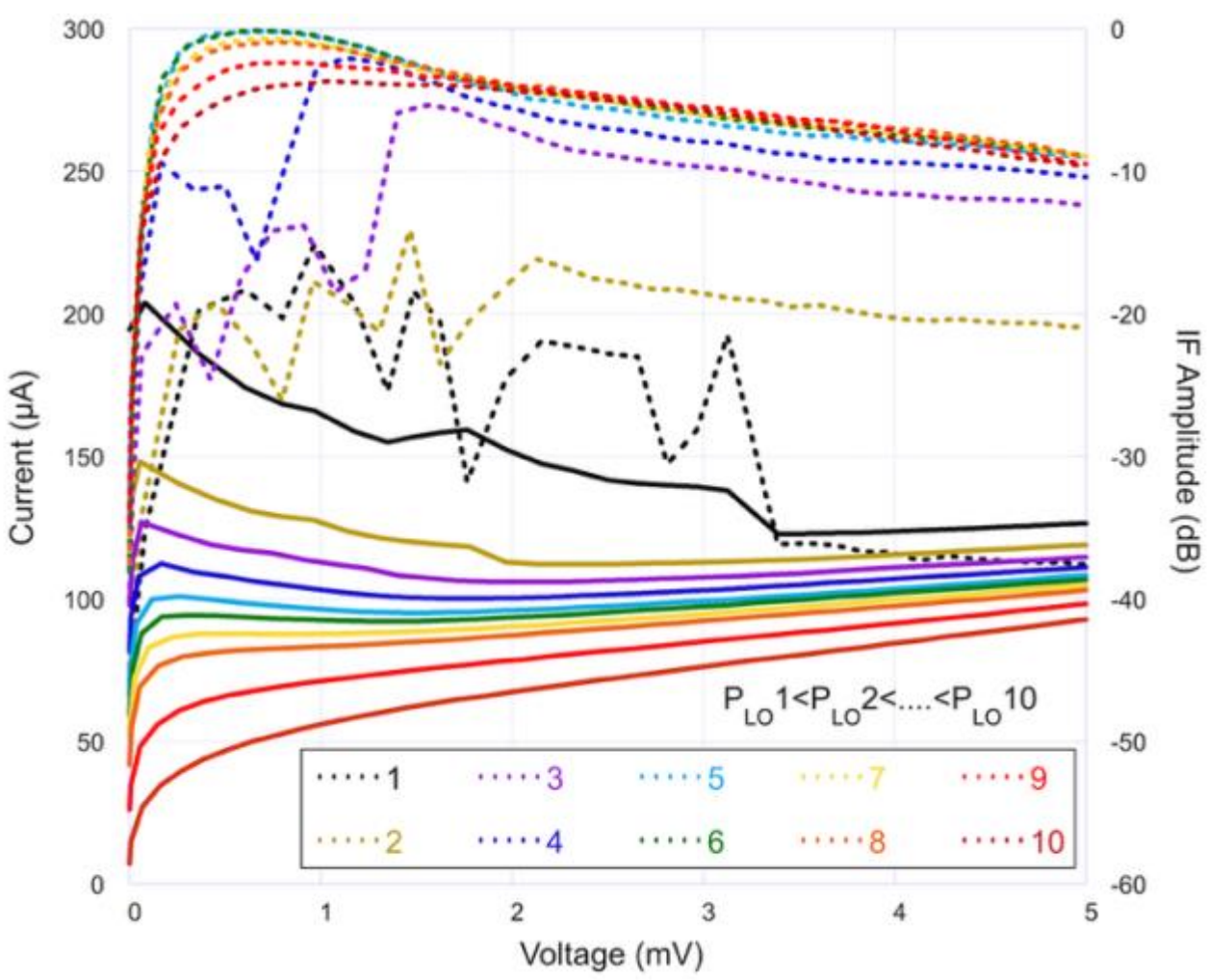

**Fig. 3.** Gradual transition from NDR to PDR as LO power increases (curves numbers from 10 to 1), and $L_B$ = 70 nH. Solid lines are IVCs, dashed lines are IF output power (proportional to the conversion gain). Same line colors correspond to the same $P_{LO}$ values. Gain curves 5 and 6 are almost indistinguishable. The dropouts in the gain vs bias curves indicate unstable operation. See [1] for explanation of the shape of the IVC under unstable conditions.
Adapted from [1], with permission from the American Institute of Physics, 2026.

## III. Gain and Bias Stability in the HEB Mixer with NDR

Within a lumped-element model [5], the HEB mixer conversion gain is:

$$\eta = 2\mathcal{L}^2 (P_{LO}/P_J) \frac{RR_L}{(R+R_L)^2} \frac{1}{\left(1+\mathcal{L}\frac{R-R_L}{R+R_L}\right)^2} \frac{1}{1+(2\pi f \tau^*)^2}, \tag{2a}$$

where

$$\tau^* = \tau_{th} / \left(1 + \mathcal{L}\frac{R-R_L}{R+R_L}\right) \tag{2b}$$

is the time constant including the ETF effect through the IF line, which defines the IF bandwidth of the mixer $\Delta f_{IF} = (2\pi\tau^*)^{-1}$.

Work [3] considered the HEB mixer conversion gain with $\mathcal{L}$ >1, primarily to highlight instability risks. Bias instabilities and parasitic oscillations have indeed been observed [6-8]. Reference [1] explained these effects from a circuit perspective. Here we extended the analysis to derive stability boundaries in the current HEB device and guide future NDR-based gain enhancement.

Extensive stability analysis have been carried out, for example, for resonant-tunneling diodes (RTDs) [9, 10]. We should also mention that a stability analysis leading to the same results was performed for transition-edge sensors (TES) [4, 11]. A simple and effective way to summarize the stability criteria is to use the complex circuit impedance. If we consider the bias circuit of Fig. 2, the following conditions should be met:

$$R_B < |Z(0)|, \tag{3a}$$
$$f_r < f_x. \tag{3b}$$

Condition (3a) is straightforward: it simply means that the bias line must intersect the IVC at only one point. Condition (3b) is more complex. It requires that the real part of the negative resistance Re$Z(f)$ becomes positive at a frequency $f_r$, which is lower than the circuit's resonant frequency $f_x$, where the imaginary part of the impedance Im$Z(f)$ becomes zero.

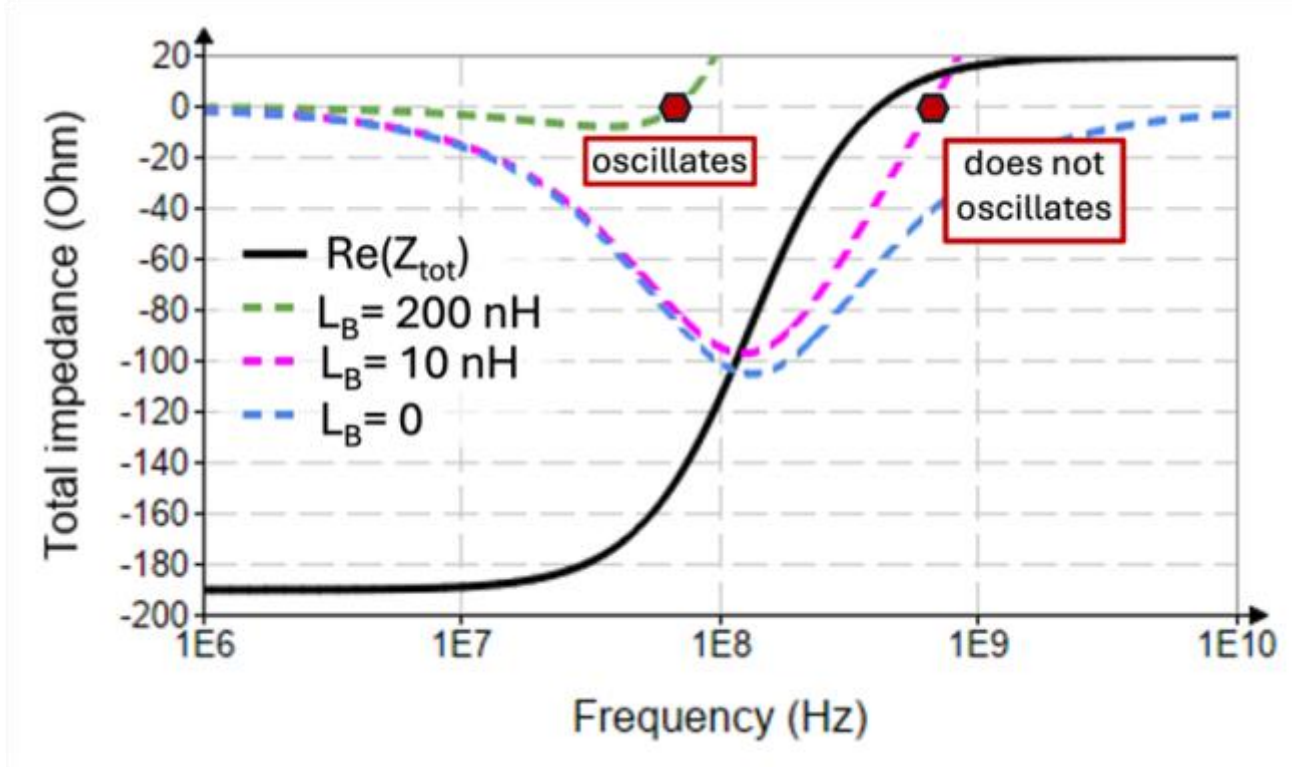

**Fig. 4.** Total impedance of the bias circuit of Fig. 2 for different values of inductance $L_B$. Soid line is Re($Z_{tot}$), dashed lines are Im($Z_{tot}$).

The frequency dependence $Z(f)$ in RTD deivces is due to parasitic reactance. However, in superconducting bolometers, the intrinsic thermal inertia creates a reactive circuit effect that can be represented by an equivalent capacitance $C_N = \tau_{th}/(2R\mathcal{L})$, where $\tau_{th}$ is the thermal time constant of the bolometer.

Strictly speaking, NbN HEB cannot be described by a single time constant as both the electro-phonon scattering ($\tau_{eph}$) and the thermal-phonon escape ($\tau_{es}$) contribute to thermal relaxation. However, in the experiment, the relaxation appears to be govern by a single time constant that depends on both the temperature and the film thickness [12]. Hence, the subsequent discussion will use the lumped-element bolometer model with $\tau_{th}$ = 60 ps (IF bandwidth $\Delta f_{IF} \approx$ 2.7 GHz) as in the current HEB device.

We consider the stability of the circuit loop including only the HEB, $L_B$, and $R_B$ (see Fig. 2). The LNA and the isolator represent a bandpass filter with the passband of ≈ 0.8-2 GHz, which is at much higher frequency than a typical value of $f_{osc}$= 10-100 MHz. Then the total impedance is:

$$Z_{tot}(f) = Z(f) + j \cdot 2\pi f L_B + R_B, \qquad (4a)$$

where

$$Z(f) = R\frac{1+\mathcal{L}}{1-\mathcal{L}} \cdot \frac{1+j\cdot 2\pi f \tau_{th}/(1+\mathcal{L})}{1+j\cdot 2\pi f \tau_{th}/(1-\mathcal{L})} \qquad (4b)$$

is the HEB impedance [1, 5].

Figure 4 illustrates condition (3b) using the $Z(f)$ plot. Here we use parameters found in our experimental device at some bias point with NDR (e.g., $\mathcal{L}$ = 1.15, $Z(0)$ ≈ -200 Ohm, etc.). In the absence of inductance, the reactance of the circuit is pure capacitive, and the $f_x$-point is simply absent. A small inductance $L_B$ = 10 nH leads to the crossing of the Im($Z_{tot}$) line with x-axis but Re($Z_{tot}$) > 0 so the oscillations are dumped. A larger inductance $L_B$ = 200 nH shifts $f_x$ towards lower frequency. Since Re($Z_{tot}(f)$) does not depend on $L_B$, then at some point, $f_r$ becomes greater than $f_x$. Hence, the oscillations become possible.

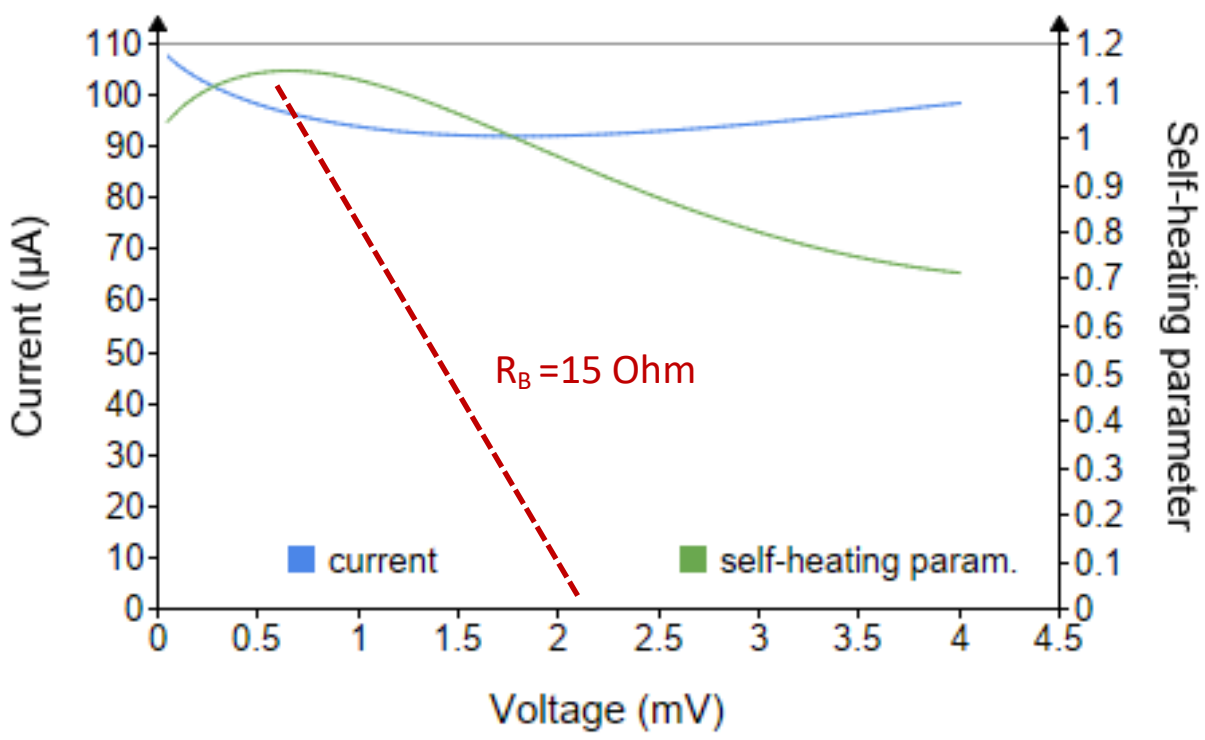

**Fig. 5.** Simulated experimental IVC (blue curve) and self-heating parameter $\mathcal{L}$ computed using (1) (green curve). The purple dashed line shows the load-line slope corresponding to the dc bias resistance $R_B$ = 15 Ohm used in the experiment.

## IV. Stability Analysis of the Experimental HEB Device

We analyze the data shown in Fig. 3. In particular, the ultimate IVC #5 with NDR, which was stable throughout the entire bias range, was used. To obtain smooth functions suitable for analysis, the experimental IVC was fitted with an analytical function from which $R$, $Z(0)$ and $\mathcal{L}$ values (using (1)) were computed as functions of the bias voltage (see Fig. 5).

Using relationships from Table I in [1], equation (3b) can be transformed into the following form:

$$\tau_{th}/(\mathcal{L} - 1) > L_B/(R + R_B). \qquad (5)$$

This condition is analogous to the stability criterion for TES devices (e.g., equation (11) in [11]). In simple terms, it means that the thermal time constant must be greater than the electrical time constant of the circuit.

Conditions (3a) and (5) are plotted in Fig. 6 as $R_B$ vs bias voltage for two values of the inductance: $L_B$ = 70 nH (as in the current experiment) and $L_B$ = 35 nH. Bias conditions corresponding to (5) are the area above the red curves, whereas (3a) is satisfied in the area below the blue curve. The voltage span (up to 1.65 mV) corresponds to the range where NDR exists.

For $L_B$ = 70 nH, stability at $V_B$ ≈ 0.9-1.0 mV (the maximum value of $\mathcal{L}$, see Fig. 5) is possible within a limited range of parameters and requires a high value of $R_B$ ≈ 150 Ohm. The experimental $R_B$ was ≈ 15 Ohm, which explains why IVC #4 (Fig. 3) was unstable in this bias range. Reducing $L_B$ to 35 nH would significantly expand the stability range, allowing the maximum $\mathcal{L}$-values to fall within it. However, larger values of $R_B$ ≈ 80-100 Ohm would still be required. This combination of $L_B$ and $R_B$ has not yet been tested.

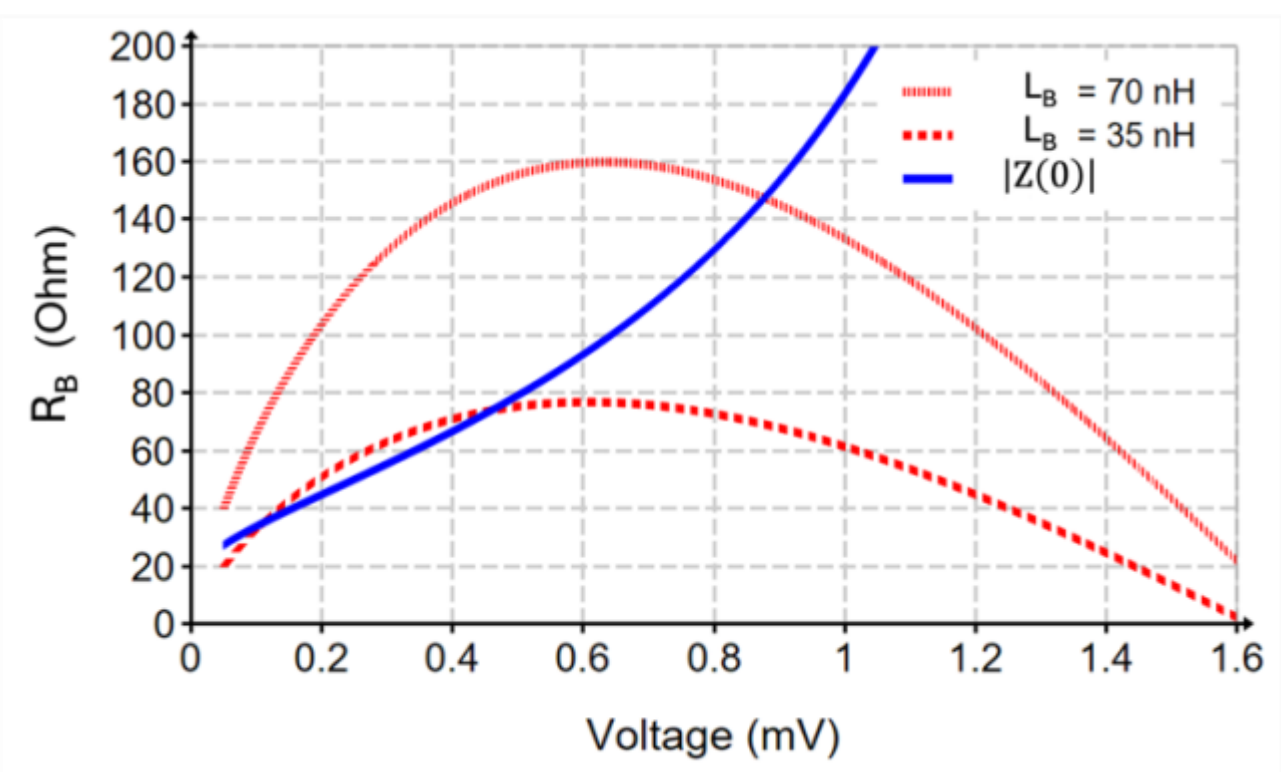

**Fig. 6.** Stability ranges for the IVC of Fig. 5. Both stability conditions (3a) and (5) are met above the red curves and below the blue curve.

Obviously, an even smaller $L_B$ value will lead to a broader stable bias range, with smaller $R_B$ values. However, the issue in this case is the loss of the IF power due to the insufficient blocking by a small inductance. For example, with $L_B$ = 35 nH, the inductive impedance at $f_0$ = 1.4 GHz is $X_L = 2\pi f_0 L_B$= 308 Ohm, which is only 6 times greater than the impedance of the IF line $R_L$ = 50 Ohm. As a result, the useful $P_{\text{IF}}$ will be reduced by ≈ 0.7 dB.

## V. Prediction of the Maximum Conversion Gain

The current experiment [1] does not provide a comprehensive outlook on the full potential of the NDR biasing. The range of parameters $R_B$ and $L_B$ was limited and this is something that needs to be addressed in future experiments. Hower, we are able to make some predictions based on the experimental data and the lumped-element model of the HEB mixer.

The model is based on the heat-balance equation for the electron temperature $T_e$:

$$\Sigma V(T_e^n - T_0^n) = P_{LO} + P_J. \quad (6)$$

Here $\Sigma$ and $n$ are material parameters. It is assumed that a single time constant $\tau_{th}(T_e) = \gamma/(n\Sigma T_e^{n-2})$ determines the thermal relaxation and that the device resistance $R(T_e)$ depends on the external factors only through their effect on the electron temperature. $\gamma$ is the Sommerfeld constant, $P_{LO}$ is the local oscillator power absorbed in the HEB device, $P_J$ is the Joule power dissipated at the bias point. This early model [2] is oversimplified in several ways and cannot yield the correct combination of bias parameters and conversion gain. However, it can simulate the relationship between $\mathcal{L}$ and $\eta$ and therefore allows prediction of the conversion gain as the IVC changes.

The resistance $R(T_e)$ is modeled with a smooth step-function simulating the superconducting transition over the temperature range $\delta T_C$ near the critical temperature $T_C$. Equation (6) is numerically solved for an array of combinations of $P_{LO}$ and $P_J$ producing a two-dimensional array of $T_e$ values. From this, all the mixer characteristics can be computed as functions of the bias and the LO power.

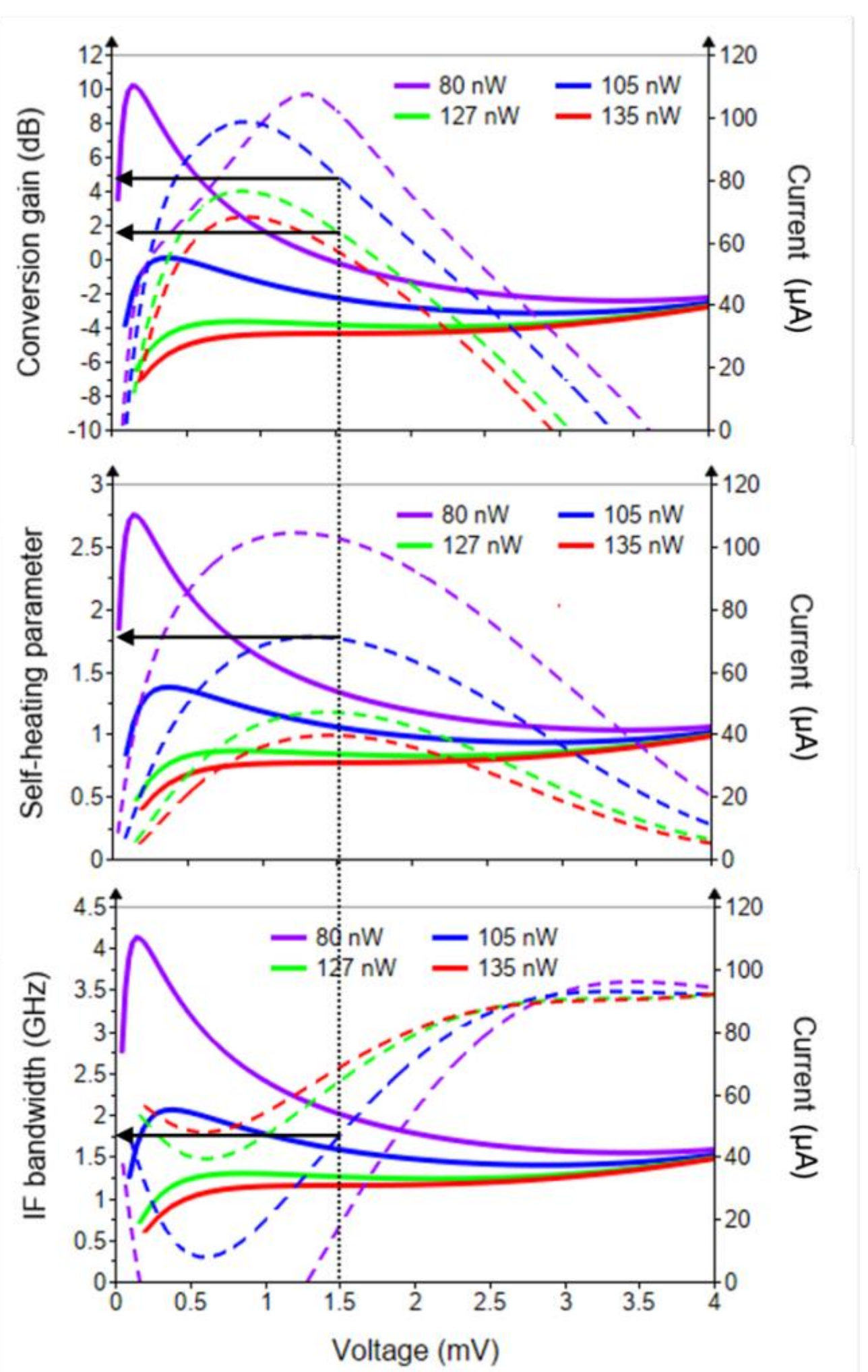

**Fig. 7.** Modeling results for conversion gain $\eta$, self-heating parameter $\mathcal{L}$, and IF bandwidth $\Delta f_{IF}$ as functions of the bias and four LO power values (shown in the legend). Solid curves are IVCs (same for each plot), dashed curves are left y-axis' variables.

Table I shows the set of NbN HEB device parameters used for modeling.

TABLE I
NbN HEB Device Parameters

| $T_C$ [K] | $\delta T_C$ [K] | $\gamma$ [JK$^{-2}$μm$^{-3}$] | $\Sigma$ [WK$^{-n}$μm$^{-3}$] | $n$ | $V$ [μm$^3$] | $R_N$ [Ohm] |
|---|---|---|---|---|---|---|
| 11.1 | 1.5 | 1.85·10$^{-16}$ | 6.2·10$^{-9}$ | 4 | 0.002 | 104 |

Here $T_C$ and $\delta T_C$ are the critical temperature and superconducting transition width, respectively. The value of $\gamma$ is taken from [13], and $n$ is from [14]. Given that $\tau_{th}$ in NbN depends on the film thickness and the acoustic transparency of the film-substrate interface, the $\Sigma$-value was adjusted so the time constant becomes $\tau_{th}(T_C)$ = 60 ps as in experiment [1].

Modeling results are presented in Fig. 7. Since NbN HEB cannot be accurately described by the lumped -element model, the simulated IVCs differ from the experimental ones. The bottom (red) IVC illustrates the case with PDR. The next (green) IVC was adjusted to yield the same maximum value of $\mathcal{L}$ =1.15 as in Fig. 5. Two upper IVC show the effect of further deceasing the $P_{LO}$. The purple curve is so steep that the modeled IF bandwidth becomes negative. Analysis of (2b) shows that this is due to the instability in the IF line where the electro-thermal feedback loop is closed through the load $R_L$ ($R_L > |Z(0)|$. This occurs at frequencies above 800 MHz (the low-frequency cut-off of the IF isolator + LNA); we do not consider this frequency range in our current bias-stability analysis, which focuses on mitigation oscillations in the 10-100 MHz range. Therefore, the approximately optimal IVC is the blue curve. Here, for $V_B \approx 1.25$ mV one can obtain simultaneously the gain ≈ 3.5-dB higher that in the current experiment and a sufficient IF bandwidth of ≈ 1.75 GHz. Note that the gain achieved in [1] with NDR bias was about 3-dB higher that the gain with PDR bias. Hence, the overall gain improvement when a smaller inductance $L_B$ is implemented may be ≈ 6 dB.

## VI. Conclusion

This work demonstrates that suppressing parasitic oscillations enables stable NDR biasing in NbN HEB mixers, significantly increasing conversion gain. Stability boundaries derived here provide guidance for future bias-circuit design, pointing toward further improvements through optimized inductance and load-line parameters.

## Acknowledgment

The authors thank B. Bumble and J. Kawamura for providing NbN HEB devices for this work, and A. Babenko for contribution and fruitful discussions on different aspect of this research.